\documentclass{jetpl}
\twocolumn
\usepackage{graphicx}
\usepackage{bm}
\lat
\usepackage[usenames,dvipsnames]{color} % for color
\newcommand{\mkrm}[1]{}           % use this to hide what has been removed

\title{Phonon-particle coupling effects in single-particle energies of semi-magic nuclei.}%

\rtitle{Phonon-particle coupling effects in single-particle energies of semi-magic nuclei.}%

\sodtitle{Phonon-particle coupling effects in single-particle energies of semi-magic nuclei.}%

\author{
 E.\,E.\,Saperstein$^{*,**}$\/\thanks{e-mail: Sapershtein\_EE@nrcki.ru},
 M.\,Baldo$^{\dagger}$,
 S.\,S.\,Pankratov$^{*,***}$,
 S.\,V.\,Tolokonnikov$^{*,***}$}

\rauthor{E.\,E.\,Saperstein, M.\,Baldo, S.\,S.\,Pankratov, S.\,V.\,Tolokonnikov}

\sodauthor{E.\,E.\,Saperstein, M.\,Baldo, S.\,S.\,Pankratov, S.\,V.\,Tolokonnikov}

\address{$^{*}$National Research Centre Kurchatov Institute, pl.
Akademika Kurchatova 1, Moscow, 123182 Russia\\
$^{**}$National Research Nuclear University MEPhI, 115409 Moscow, Russia\\
$^{\dagger}$Istitute Nazionale di Fisica Nucleare, Sezione di Catania, 64 Via S.-Sofia, I-95125 Catania, Italy\\
$^{***}$Moscow Institute of Physics and Technology, 141700 Dolgoprudny, Russia}

\dates{\today}{*}

\abstract{ A method is presented to evaluate the particle-phonon coupling (PC) corrections to the
single-particle energies (SPEs) in  semi-magic nuclei. In such nuclei always there is a collective
low-lying $2^+$ phonon, and a strong mixture of single-particle and particle-phonon states often
occurs. As in magic nuclei, the so-called $g^2_L$ approximation, where $g_L$ is the vertex of the
$L$-phonon creation, can be used for finding the PC correction $\delta \Sigma^{\rm PC}(\eps)$ to the
initial mass operator $\Sigma_0$.  In addition to the usual pole diagram, the phonon ``tadpole''
diagram is also taken into account. In semi-magic nuclei, the perturbation theory in $\delta
\Sigma^{\rm PC}(\eps)$ with respect to $\Sigma_0$ is often invalid for finding the PC corrected SPEs.
Instead, the Dyson equation with the mass operator $\Sigma(\eps){=}\Sigma_0{+}\delta \Sigma^{\rm
PC}(\eps)$ is solved directly, without any use of the perturbation theory. Results for a chain of
semi-magic Pb isotopes are presented.}


\begin{document}

\newcommand{\beq}{\begin{equation}}
\newcommand{\eeq}{\end{equation}}
\newcommand{\bea}{\begin{eqnarray}}
\newcommand{\eea}{\end{eqnarray}}
\newcommand{\eps}{\varepsilon}
\newcommand{\bfg}{\boldsymbol}

\maketitle

Last decade, there was a revival of the interest within different self-consistent nuclear approaches
to study the particle-phonon coupling (PC) effects in the single-particle energies (SPEs) of magic
nuclei. We cite here such studies within the relativistic mean-field theory \cite{Litv-Ring}, within
the Skyrme--Hartree--Fock method \cite{Bort,Dobaczewski,Baldo-PC} and on the basis of the
self-consistent theory of finite Fermi systems (TFFS) \cite{levels}. The Fayans energy density
functional (EDF) was used in the last case to find the self-consistent basis.    In all the references
cited above double-magic nuclei were considered. There are several reasons for such choice. First,
these nuclei are non-superfluid which simplifies the theoretical analysis. Second, the so-called
$g^2_L$ approximation is, as a rule, valid in magic nuclei,  $g_L$ being the vertex of the $L$-phonon
creation. Moreover, the perturbation theory in terms of the PC correction to the mass operator ia also
applicable, which makes evaluation of PC corrected SPEs rather simple. At last, there is a lot of
experimental data on SPEs in these nuclei \cite{exp}.

In this work, we extend the field of this problem to semi-magic nuclei. Unfortunately, the
experimental data on the SPEs in semi-magic nuclei are rather scarce. Indeed, the single-particle
spectroscopic factor $S$ of an excited state under consideration should be known. In addition, its
value should be rather large, in order that one can interpret this state as a single-particle one.
Extraction of the spectroscopic factors from the reaction data is a complicated theoretical problem,
therefore the list of known spectroscopic factors is rather limited. However, the PC corrections to
the SPEs are necessary not only by themselves, they are also usually important ingredients of the
procedure of finding PC corrections to other nuclear characteristics, e.g., magnetic moments and $M1$
transitions in odd nuclei \cite{PC-mu1,PC-mu2,PC-mu3}. PC corrections to the double odd-even mass
differences found in the approach starting from the free $NN$ potential \cite{DMD1,DMD2} is another
example where the PC corrections to SPEs are of primary importance.

A semi-magic nucleus consists of two sub-systems with different properties. One of them, magic, is
normal, whereas the non-magic counterpart is superfluid. We will consider the SPEs of the normal
sub-system only. Therefore, the main part of the formalism for description of the PC corrections
developed for magic nuclei \cite{levels} remains valid. One difference with double-magic nuclei is
that the vertex $g_L({\bf r})$ obeys the QRPA-like TFFS equation in superfluid nuclei, in contrast to
 simple  RPA-like equation in magic nuclei. This complication is not very  serious as we developed
the necessary method with the use of the Fayans EDF in a previous paper  \cite{BE2}. A real difficulty
arises in non-magic nuclei due to appearance of low-lying $2^+$ states which is a characteristic
feature of such nuclei. As a result, small denominators appear regularly in the expressions for the PC
corrections which makes unapplicable a plane perturbation theory.

To find the SPEs with account for the PC effects, we solve the following equation: \beq \left(\eps-H_0
-\delta \Sigma^{\rm PC}(\eps) \right) \phi =0, \label{sp-eq}\eeq where $H_0$ is the quasiparticle
Hamiltonian with the spectrum $\eps_{\lambda}^{(0)}$ and $\delta \Sigma^{\rm PC}$ is the PC correction
to the quasiparticle mass operator. This is equivalent to the Dyson equation for the one-particle
Green function $G$ with the mass operator $\Sigma=\Sigma_0+\delta \Sigma^{\rm PC}(\eps)$.

In magic nuclei \cite{levels,DMD1,DMD2} the perturbation theory in $\delta \Sigma^{\rm PC}$ with
respect to $H_0$ was used to solve this equation: \beq \eps_{\lambda}=\eps_{\lambda}^{(0)} +
Z_{\lambda}^{\rm PC} \delta \Sigma^{\rm PC}_{\lambda\lambda}(\eps_{\lambda}^{(0)})
,\label{eps-PC0}\eeq where \beq Z_{\lambda}^{\rm PC} =\left({1- \left(\frac {\partial} {\partial \eps}
 \delta \Sigma^{\rm PC}(\eps) \right)_{\eps=\eps_{\lambda}^{(0)}}}\right)^{-1}. \label{Z-fac}\eeq

 In this article, we will solve Eq. (\ref{sp-eq}) directly, without any additional approximations.
As to the $g_L^2$-approximation for the PC correction $\delta \Sigma^{\rm PC}$ to the mass operator,
it remains valid in semi-magic nuclei, and only the next step from (\ref{sp-eq}) to (\ref{eps-PC0})
becomes invalid. In the case when several $L$-phonons are taken into account, the total PC variation
of the mass operator in Eqs. (\ref{sp-eq})--(\ref{Z-fac}) is just the sum: \beq \delta \Sigma^{\rm PC}
= \sum_L \delta \Sigma^{\rm PC}_L . \label{sum-L}\eeq

The diagrams for the $\delta \Sigma^{\rm PC}_L$ operator within the $g_L^2$-approximation are
displayed in Fig. 1. The first one is the usual pole diagram, with obvious notation, whereas the
second, ``tadpole'' diagram represents the sum of all non-pole diagrams of order $g_L^2$.

\begin{figure}
\centerline {\includegraphics [width=80mm]{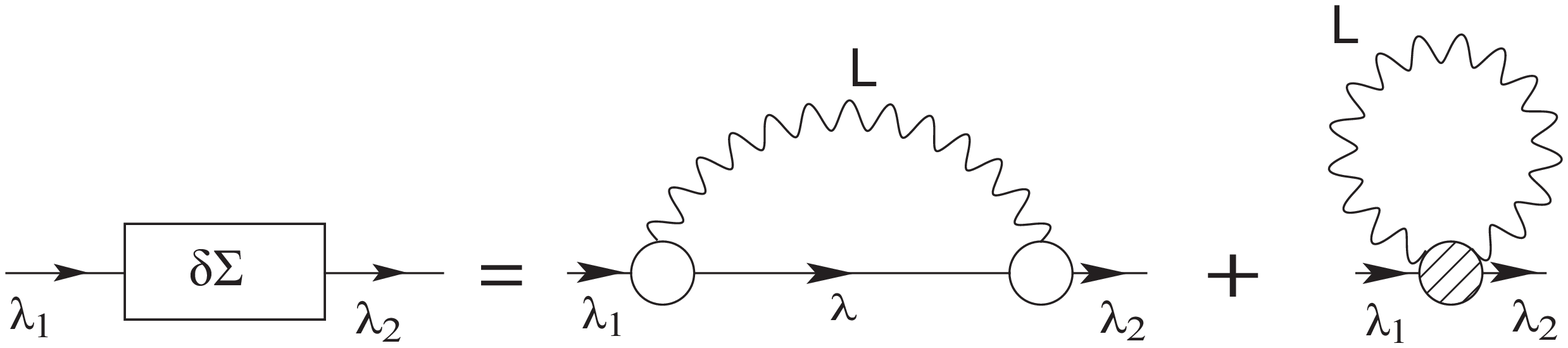}} \caption{Fig. 1. PC corrections to the mass
operator. The gray blob denotes the phonon ``tadpole'' term.} \label{fig:SigPC}
\end{figure}

Explicit expression for the pole term is well known \cite{Litv-Ring,levels,DMD1}, and we present it
just for completeness: \bea \delta\Sigma^{\rm pole}_{\lambda\lambda}(\epsilon)&=&\sum_{\lambda_1\,M}
|\langle\lambda_1|g_{LM}|\lambda\rangle|^2 \nonumber\\
&\times&\left(\frac{n_{\lambda_1}}{\eps+\omega_L-
\eps_{\lambda_1}^{(0)}}+\frac{1-n_{\lambda_1}}{\eps-\omega_L -\eps_{\lambda_1}^{(0)}}\right),
\label{dSig2} \eea where $\omega_L$ is the excitation energy of the $L$-phonon and
$n_{\lambda}{=}(1,0)$ are the particle occupation number (remind that we deal with the normal
subsystem of a semi-magic nucleus).

The vertex $g_L(\bf r)$ obeys the QRPA-like TFFS equation \cite{AB}, \beq {\hat g}_L(\omega)=
 {\hat {\cal F}} {\hat A}(\omega) {\hat g}_L(\omega), \label{gL} \eeq
where all the terms are  matrices. The angular momentum projection $M$, which is written down in Eq.
(\ref{dSig2}) explicitly, is here and below for brevity omitted. In the standard TFFS notation, we
have: \beq {\hat g}_L=\left(\begin{array}{c}g_L^{(0)}
\\g_L^{(1)}\\g_L^{(2)}\end{array}\right),
\label{gs} \eeq \beq {\hat {\cal F}}=\left(\begin{array}{ccc}
{\cal F} &{\cal F}^{\omega \xi}&{\cal F}^{\omega \xi}\\
{\cal F}^{\xi \omega }&{\cal F}^\xi  &{\cal F}^{\xi \omega }\\
{\cal F}^{\xi \omega }&{\cal F}^{\xi \omega }& {\cal F}^\xi \end{array}\right). \label{Fs} \eeq

In (\ref{gs}), $g^{(0)}$ is the normal component of the vertex ${\hat g}$, whereas \mkrm{as}  $g^{(1),(2)}$
are two anomalous ones. In Eq. (\ref{Fs}), $\cal{ F}$ is the usual Landau--Migdal interaction amplitude
which is the second variation derivative of the  EDF ${\cal E} [\rho,\nu]$ over the normal density
$\rho$ . The effective pairing interaction $\cal{ F}^\xi$ is the second derivative of the EDF over the
anomalous density $\nu$. At last, the amplitude ${\cal F}^{\xi \omega }$ stands for the mixed
derivative of ${\cal E}$ over $\rho$ and $\nu$.

The matrix ${\hat A}$ consists of $3\times 3$  integrals over $\eps$ of the products of different
combinations of the Green function $G(\eps)$ and two Gor'kov functios $F^{(1)}(\eps)$ and
$F^{(2)}(\eps)$  \cite{AB}.

As we need the proton vertex ${\hat g}_L^p$ and the proton subsystem  is normal, only the normal
vertex $g_L^{(0)p}$ is non-zero in this case.  This is explicit meaning of the short notation $g_L$ in
(\ref{dSig2}) and below.

  For solving the above equations, we use the self-consistent basis generated by the version DF3-a
\cite{DF3-a} of the Fayans EDF \cite{Fay1,Fay}. The nuclear mean-field potential $U(r)$ is the first
derivative of  ${\cal E}$ over $\rho$.

All low-lying phonons we deal with  are of surface nature,  the surface peak dominating in their
creation amplitude: \beq g_L(r)=\alpha_L \frac {dU} {dr} +\chi_L(r). \label{gLonr}\eeq The first term
in this expression  is surface peaked, whereas the in-volume term $\chi_L(r)$ is rather small. It is
illustrated in Fig. 2 for the $2^+_1$ and $3^-_1$ states in $^{204}$Pb. In this work, just as in
\cite{levels,DMD1,DMD2}, we neglect the in-volume term in (\ref{gLonr}) when considering the tadpole
PC term of $\delta \Sigma_L^{\rm PC}$.  In the result, it is reduced, see Ref. \cite{levels}, to
rather simple form: \beq \delta\Sigma^{\rm tad}_L = \frac {\alpha_L ^2} 2 \frac {2L+1} 3 \triangle
U(r). \label{tad-L}\eeq

\begin{figure}
\centerline {\includegraphics [width=80mm]{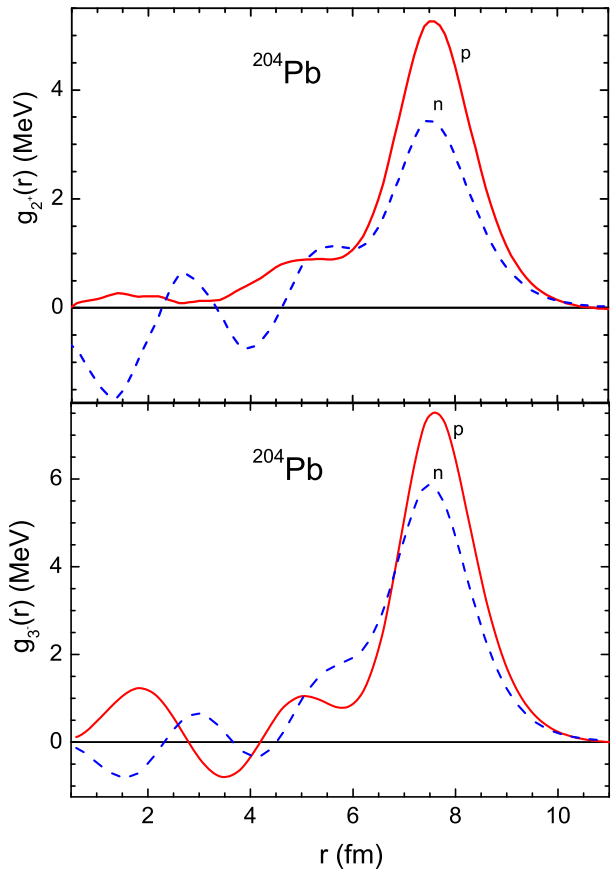}} \caption{Fig2. (Color online) Phonon creation
amplitudes $g_L(r)$ for two low-lying phonons in $^{204}$Pb.} \label{fig:gL}
\end{figure}

\begin{figure}
\centerline {\includegraphics [width=80mm]{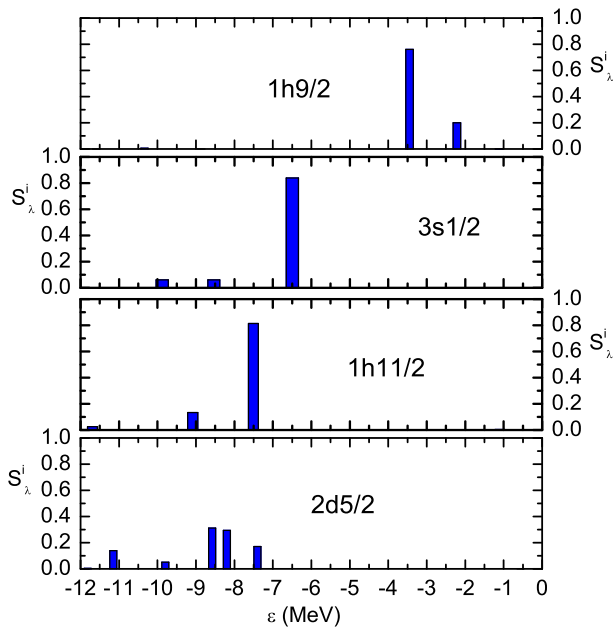}} \caption{Fig3. (Color online)  $S$-factors.}
\label{fig:Slam}
\end{figure}

Note that the above scheme for the ghost ($L{=}1,\;\omega_1{=}0$) phonon results in an explicit
expression for the ``recoil effect.'' Details can be found in \cite{levels}.

In this work, we limit ourselves with four even lead isotopes, $^{200,202,204,206}$Pb. In all cases we
consider two low-lying phonons, $2^+_1$ and $3^-_1$. Their excitation energies are presented in Table
1. As one can see, they agree with existing experimental data sufficiently well. The ghost $1^-$ is
also taken into account, although the corresponding correction for nuclei under consideration is very
small, because it depends on the mass number as $1/A$.

\begin{table}[h!]
\caption{Table 1. Excitation energies  $\omega_{2,3}$ (MeV) of the $2^+_1$ and $3^-_1$ phonons in even
Pb isotopes.}
\begin{tabular}{c| c| c| c| c  }

\hline\noalign{\smallskip} nucleus  & $\omega_2^{\rm th}$   & $\omega_2^{\rm exp}$
& $\omega_3^{\rm th}$     & $\omega_3^{\rm exp}$    \\
\noalign{\smallskip}\hline\noalign{\smallskip}
$^{200}$Pb& 0.789      & 1.026 & 2.620    &  -      \\
$^{202}$Pb& 0.823      & 0.960 & 2.704    & 2.517  \\
$^{204}$Pb& 0.882      & 0.899 & 2.785    & 2.621   \\
$^{206}$Pb& 0.945      & 0.803 & 2.839    & 2.648    \\

\noalign{\smallskip}\hline

\end{tabular}\label{tab1}
\end{table}

Let us go to the  description of the method we use to solve Eq. (\ref{sp-eq}). As the PC corrections
are important only for the SPEs nearby the Fermi surface, we limit ourselves with a model space $S_0$
including two shells close to it, i.e. one hole and one particle shells, and besides we retain only
the negative energy states. To avoid any misunderstanding, we stress that this restriction concerns
only Eq. (\ref{sp-eq}). In Eq. (\ref{dSig2}) for $\delta\Sigma^{\rm pole}$, we use rather wide
single-particle space with energies $\eps_{\lambda}^{(0)}{<}40\;$MeV. We take as an example to
illustrate the method the nucleus $^{204}$Pb which is sufficiently distant from the double-magic
$^{208}$Pb to be considered as a typical semi-magic. The space $S_0$ involves 5 hole states
(1g$_{7/2}$, 2d$_{5/2}$, 1h$_{11/2}$, 2d$_{3/2}$, 3s$_{1/2}$) and four particle ones (1g$_{9/2}$,
2f$_{7/2}$, 1i$_{13/2}$, 2f$_{5/2}$). We see that there is here only one state for each $(l,j)$ value.
Therefore, we need  only diagonal elements $\delta\Sigma^{\rm pole}_{\lambda \lambda}$ (\ref{dSig2}),
which  simplifies very much  the solution of the Dyson equation.   In the result, Eq. (\ref{sp-eq})
reduces as follows: \beq \eps-\eps_{\lambda}^{(0)}-\delta \Sigma^{\rm PC}_{\lambda \lambda}(\eps) =0,
\label{sp-eq1}\eeq

The tadpole term does not depend on the energy, therefore only poles of Eq. (\ref{dSig2}) are the
singular points of Eq. (\ref{sp-eq1}). They can be readily found from (\ref{dSig2}) in terms of
$\eps_{\lambda}^{(0)}$ and $\omega_L$. It can be easily seen that the lhs of Eq. (\ref{sp-eq1}) always
changes sign between any couple of neighboring poles, and the corresponding solution
$\eps_{\lambda}^i$ can be found with usual methods. In this notation, $\lambda$ is just the index for
the initial single-particle state from which the state $|\lambda,i\rangle$ originated. The latter is a
mixture of a single-particle state with several particle-phonon states. The corresponding
single-particle strength distribution factors ($S$-factors) can be found similar to (\ref{Z-fac}):
\beq S_{\lambda}^i =\left(1- \left(\frac {\partial} {\partial \eps}
 \delta \Sigma^{\rm PC}(\eps) \right)_{\eps=\eps_{\lambda}^{i}}\right)^{-1}. \label{S-fac}\eeq
 Evidently, they should obey the normalization rule: \beq \sum_i S_{\lambda}^i =1.   \label{norm}\eeq
Accuracy of fulfillment of this relation is a measure of validity of the model space $S_0$ we use to
solve the problem under consideration. In the major part of the cases we will consider it is fulfilled
with 1--2\% accuracy.

A set of solutions for four $|\lambda,i\rangle$ states in $^{204}$Pb is presented in Table 2. The
corresponding $S$-factors are displayed in Fig. 3. In three upper cases for a given $\lambda$ there is
a state $|\lambda,i_0\rangle$ with dominating $S_{\lambda}^{i_0}$ value (${\simeq}0.8$). They are
examples of ``good'' single-particle states. In such cases, the following prescription looks natural
for the PC corrected single-particle characteristics: \beq \eps_{\lambda}={\eps}_{\lambda}^{i_0}; \;
Z_{\lambda}^{\rm PC}=S_{\lambda}^{i_0}. \label{i0} \eeq This is an analog of Eqs. (\ref{eps-PC0}) and
(\ref{Z-fac}) in the perturbative  solution.

\begin{table}[]
\caption{Table 2. Examples of solutions of Eq. (\ref{sp-eq1}) for protons in $^{204}$Pb.}
\begin{tabular}{ c| c| c | l }

\hline\noalign{\smallskip}

$\lambda$   & $i$ & $\eps_{\lambda}^i$  &\hspace*{7mm} $S_{\lambda}^i$   \\

\noalign{\smallskip}\hline\noalign{\smallskip}

2$d_{5/2}$ &  1  &   -11.817 &  0.314 $\times 10^{-2}$ \\
           &  2  &   -11.150 &  0.139 \\
           &  3  &    -9.799 &  0.516 $\times 10^{-1}$ \\
           &  4  &    -8.580 &  0.312 \\
           &  5  &    -8.195 &  0.295 \\
           &  6  &    -7.404 &  0.171 \\
           &  7  &    -0.564 &  0.791 $\times 10^{-3}$ \\
\noalign{\smallskip}\hline\noalign{\smallskip}

1$h_{11/2}$ &  1 &    -13.717  &  0.692 $\times 10^{-3}$ \\
            &  2 &    -11.690  &  0.256 $\times 10^{-1}$ \\
            &  3 &     -9.084  &  0.134 \\
            &  4 &     -7.509  &  0.814 \\
            &  5 &     -2.471  &  0.427 $\times 10^{-3}$ \\
            &  6 &     -1.095  &  0.473 $\times 10^{-2}$ \\
\noalign{\smallskip}\hline\noalign{\smallskip}

3$s_{1/2}$  &  1  &    -9.877  &  0.608 $\times 10^{-1}$ \\
            &  2  &    -8.536  &  0.604 $\times 10^{-1}$ \\
            &  3  &    -6.493  &  0.839 \\
\noalign{\smallskip}\hline\noalign{\smallskip}

1$h_{9/2}$ &  1  &   -13.736  &  0.220 $\times 10^{-2}$ \\
           &  2  &   -11.596  &  0.777 $\times 10^{-3}$ \\
           &  3  &   -10.339  &  0.674 $\times 10^{-2}$ \\
           &  4  &    -8.862  &  0.484 $\times 10^{-3}$ \\
           &  5  &    -3.447  &  0.760 \\
           &  6  &    -2.217  &  0.199 \\
           &  7  &    -1.122  &  0.288 $\times 10^{-2}$ \\

 \noalign{\smallskip}\hline

\end{tabular}\label{tab2}
\end{table}

\begin{table}[h!]
\caption{Table 3. PC corrected proton single-particle characteristics $\eps_\lambda$ and $Z_\lambda$
of even Pb isotopes. The total correction to the SPE $\delta \eps_{\lambda}^{\rm
PC}{=}\eps_{\lambda}{-}\eps_{\lambda}^{(0)}$ is presented. The corresponding tadpole correction
$\delta \eps_\lambda^{\rm tad}$ from Eq. (\ref{tad-L})  is given separately.}
\begin{tabular}{c|c|c|c|c|c|c}
\hline

 \!\!Nucleus\!\!& $\lambda$ & $\eps^{(0)}_\lambda$ & $\delta \eps_\lambda^{\rm tad}$ & $\delta \eps_\lambda^{\rm
PC}$ &$\eps_\lambda$ & $Z_\lambda$ \\
\hline

\!\!$^{200}$Pb& 1i$_{13/2}$  &   -0.26   &   0.39   &   0.13   &   -0.13   &   0.96   \\ % calc
          & 2f$_{7/2}$   &   -1.05   &   0.24   &  -0.30   &   -1.35   &   0.83   \\ % calc
          & 1h$_{9/2}$   &   -2.33   &   0.33   &   0.12   &   -2.21   &   0.93   \\ % calc
          & 3s$_{1/2}$   &   -5.81   &   0.20   &   0.01   &   -5.80   &   0.89   \\ % calc
          & 2d$_{3/2}$   &   -6.67   &   0.21   &   0.17   &   -6.50   &   0.89   \\ % calc
          & 1h$_{11/2}$  &   -7.06   &   0.37   &   0.25   &   -6.81   &   0.93   \\ % calc
          & 2d$_{5/2}$   &   -7.88   &   0.21   &   0.28   &   -7.60   &   0.88   \\ % calc
          & 1g$_{7/2}$   &   -9.97   &   0.29   &   0.08   &   -9.89   &   0.91   \\ % calc

\hline

\!\!$^{202}$Pb& 1i$_{13/2}$  &   -0.74   &   0.41   &   0.13   &   -0.61   &   0.95  \\ % calc
          & 2f$_{7/2}$   &   -1.52   &   0.25   &  -0.29   &   -1.81   &   0.83  \\ % calc
          & 1h$_{9/2}$   &   -2.86   &   0.34   &   0.13   &   -2.73   &   0.93  \\ % calc
          & 3s$_{1/2}$   &   -6.26   &   0.21   &   0.01   &   -6.25   &   0.89  \\ % calc
          & 2d$_{3/2}$   &   -7.09   &   0.22   &   0.17   &   -6.92   &   0.89  \\ % calc
          & 1h$_{11/2}$  &   -7.52   &   0.38   &   0.25   &   -7.27   &   0.93  \\ % calc
          & 2d$_{5/2}$   &   -8.34   &   0.22   &   0.30   &   -8.04   &   0.87  \\ % calc
          & 1g$_{7/2}$   &  -10.46   &   0.30   &   0.08   &  -10.38   &   0.91  \\ % calc

\hline

\!\!$^{204}$Pb& 1i$_{13/2}$  &   -1.21   &   0.32   &   0.14   &   -1.07   &   0.97  \\ % calc
          & 2f$_{7/2}$   &   -2.01   &   0.20   &  -0.23   &   -2.24   &   0.87  \\ % calc
          & 1h$_{9/2}$   &   -3.36   &   0.27   &   0.17   &   -3.19   &   0.96  \\ % calc
          & 3s$_{1/2}$   &   -6.72   &   0.17   &   0.23   &   -6.49   &   0.84  \\ % ok
          & 2d$_{3/2}$   &   -7.51   &   0.17   &   0.05   &   -7.46   &   0.94  \\ % calc
          & 1h$_{11/2}$  &   -7.98   &   0.30   &   0.25   &   -7.73   &   0.95  \\ % calc
          & 2d$_{5/2}$   &   -8.80   &   0.18   &   0.17   &   -8.63   &   0.92  \\ % calc
          & 1g$_{7/2}$   &  -10.93   &   0.24   &   0.13   &  -10.80   &   0.95  \\ % calc

\hline

\!\!$^{206}$Pb& 1i$_{13/2}$  &   -1.67   &   0.30   &   0.07   &   -1.60   &   0.89  \\ % calc
          & 2f$_{7/2}$   &   -2.51   &   0.19   &  -0.30   &   -2.81   &   0.82  \\ % calc
          & 1h$_{9/2}$   &   -3.82   &   0.25   &   0.19   &   -3.63   &   0.97  \\ % calc
          & 3s$_{1/2}$   &   -7.18   &   0.16   &   0.08   &   -7.10   &   0.89  \\ % ok
          & 2d$_{3/2}$   &   -7.91   &   0.16   &   0.11   &   -7.80   &   0.86  \\ % calc
          & 1h$_{11/2}$  &   -8.42   &   0.27   &   0.37   &   -8.05   &   0.88  \\ % ok
          & 2d$_{5/2}$   &   -9.28   &   0.16   &   0.12   &   -9.16   &   0.95  \\ % calc
          & 1g$_{7/2}$   &  -11.36   &   0.22   &   0.24   &  -11.12   &   0.90  \\ % calc
\hline
\end{tabular}\label{tab3}
\end{table}

The lowest panel in Fig. 3 represents a case of a strong spread where there are two or more numbers
$i$ with comparable values of the spectroscopic factors $S_{\lambda}^i$. In such cases, we suggest the
following generalization of Eq. (\ref{i0}): \beq  \eps_{\lambda}= \frac 1 {Z_{\lambda}^{\rm PC}}\sum_i
{\eps}_{\lambda}^i S_{\lambda}^i,\label{spread1}\eeq where \beq Z_{\lambda}^{\rm PC}=\sum_i
S_{\lambda}^i. \label{spread2}\eeq

\begin{figure}
\centerline {\includegraphics [width=80mm]{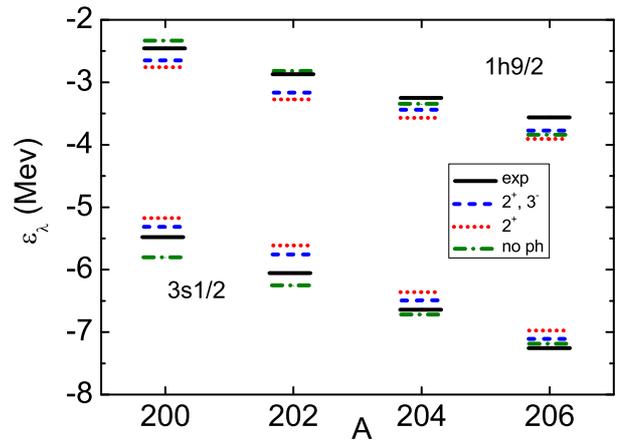}} \caption{Fig4. (Color online) SPEs corresponding to
the ground states of the odd Tl and Bi isotopes, the proton-odd neighbors of the $^A$Pb nucleus, the
$3s_{1/2}$ and $1h_{9/}2$ states correspondingly, with and without PC corrections.} \label{fig:SPE1}
\end{figure}

In both the above sums, only  the states $|\lambda,i\rangle$ with appreciable values of
$S_{\lambda}^i$ are included. In practice, we include in these sums the states with
$S_{\lambda}^i{>}0.1$. The value of $\eps_{\lambda}$ is just the centroid of the single particle
energy distribution.

The results for SPEs and $Z$-factors are presented in Table 3. For each segment of the table which
relates to the even core $^A$Pb, the lower part, till the $3s_{1/2}$ state, describes the proton
holes, i. e. the states of the $^{A-1}$Tl nucleus. On the contrary, the upper states are the proton
particle states, i.e. they belong to the $^{A+1}$Bi nucleus. The tadpole term is given separately. The
recoil correction is not presented explicitly as it is very small in the lead region, the maximum
value is $\delta \eps_{\lambda}(1^-){=}0.04$ MeV. Thus, we may approximately include it to the pole
correction $\delta \eps_{\lambda}^{\rm pole}{=}\delta \eps_{\lambda}^{\rm PC}{-}\delta
\eps_{\lambda}^{\rm tad}$. We see that the tadpole correction is always positive, whereas the pole one
is, as a rule, negative. Their absolute values are of comparable magnitude.   What is more, the
tadpole term often dominates. In such cases, neglect with the tadpole term leads to a non-correct sign
of the total SPE correction. In other cases, calculations of the PC corrections to SPEs without the
tadpole term lead to correct signs of $\delta \eps_\lambda^{\rm PC}$  but overestimate their absolute
values significantly. Thus, for the semi-magic nuclei we see the tendency noted before in magic nuclei
\cite{levels}.

In Fig. 4, we display the SPEs corresponding to the ground states of the corresponding nuclei. In this
case, the experimental values are known as they can be found in terms of mass values of neighboring
odd and even nuclei \cite{mass}. Explicitly, they are equal to $-S_p$, the proton separation energies
taken with the opposite sign. The results are compared of two sets of calculations. In the first case,
the single $2^+_1$ phonon is taken into account, whereas both the phonons, $2^+_1$ and $3^-_1$,
participate in the second set of calculations. We see, firstly, that the role of both \mkrm{the}
phonons is comparable. Secondly, the total PC correction is rather small, and, finally, often it makes
the agreement with the data worse. This is not strange as we deal with the EDF method containing
phenomenological  parameters fitted mainly to nuclear masses, which determine the experimental SPEs
under discussion.

In general, the problem of explicit consideration of the PC corrections within the EDF method or any
other self-consistent approach operating with phenomenological parameters is rather delicate. Indeed,
these parameters include different PC effects implicitly. Therefore, a regular inclusion of the PC
corrections inevitably should be accompanied with a readjustment of the initial parameters. In such a
situation, it is a more promising method  to separate  the fluctuating part of the PC corrections,
which changes in a non-regular way from a nucleus to another. Such a strategy was chosen, e.g., in
\cite{Ca-rad} to explain an anomalous $A$ dependence of charge radii of heavy calcium isotopes found
recently by the ISOLDE collaboration \cite{Ca-rad-exp}.

A different situation occurs typically for  the  approaches  which start from the free $NN$
interactions. In these cases, the PC induced corrections should be just added to the main terms found
with a free $NN$ potential, and the PC correction to the SPEs is one of the necessary ingredients for
 such calculations. Such approach was rather popular  in the last
decade for  the nuclear pairing problem \cite{milan-1,milan-2,Pankrat-1,Pankrat-2}. An analogous
method was developed in \cite{Gnezd-1,Gnezd-2} to find the double odd-even mass differences of magic
nuclei or semi-magic ones, for the normal subsystems in the last case.  In this problem, a method to
find the PC corrections was developed for magic nuclei in \cite{DMD1,DMD2}. In that case, a plain
perturbation theory was used to find the PC corrections to the SPEs.

To resume, a method is developed to find the PC corrections to SPEs for semi-magic nuclei beyond the
perturbation theory in the PC correction to the mass operator $\delta \Sigma^{\rm PC}(\eps)$ with
respect to $\Sigma_0$. Instead, the Dyson equation with the mass operator
$\Sigma(\eps){=}\Sigma_0{+}\delta \Sigma^{\rm PC}(\eps)$ is solved directly, without any use of the
perturbation theory. The method is checked for a chain of even Pb isotopes. This makes it possible to
extend  to semi-magic nuclei the field of consistent consideration of the PC corrections to the double
odd-even mass differences and some another problems. For semi-magic nuclei under consideration, the
tadpole correction to the SPEs turned out to be of primary importance.

\vskip 0.3 cm We acknowledge for support the Russian Science Foundation, Grants Nos. 16-12-10155 and
16-12-10161. The work was also partly supported  by the Russian Foundation for Basic Research Grants 14-02-00107-a, 14-22-03040-ofi\_m
and 16-02-00228-a. Calculations were partially carried out on the Computer Center of Kurchatov
Institute.
{}

\end{document}